\documentclass[pre,aps,preprint,showpacs,preprintnumbers,amsmath,amssymb,secnumroman,eqsecnum]{revtex4}

\usepackage{graphicx}
\usepackage{dcolumn}
\usepackage{bm}

\newcommand{\beq}{\begin{equation}}
\newcommand{\eeq}{\end{equation}}
\newcommand{\beqa}{\begin{eqnarray}}
\newcommand{\eeqa}{\end{eqnarray}}
\newcommand{\vc}[1]{\mbox{\boldmath $#1$}}

\newcommand{\vol}[1]{{\bf #1}}

\newcommand{\du}[1]{{\bf\sf #1}}

\begin{document}


\title{Effect of inertia on laminar swimming and flying of an assembly of rigid spheres in an incompressible viscous fluid}

\author{B. U. Felderhof}

 \email{ufelder@physik.rwth-aachen.de}
\affiliation{Institut f\"ur Theorie der Statistischen Physik \\ RWTH Aachen University\\
Templergraben 55\\52056 Aachen\\ Germany\\
}%



\date{\today}

\begin{abstract}
A mechanical model of swimming and flying in an incompressible viscous fluid in the absence of gravity is studied on the basis of assumed equations of motion. The system is modeled as an assembly of rigid spheres subject to elastic direct interactions and to periodic actuating forces which sum to zero. Hydrodynamic interactions are taken into account in the virtual mass matrix and in the friction matrix of the assembly. An equation of motion is derived for the velocity of the geometric center of the assembly. The mean power is calculated as the mean rate of dissipation. The full range of viscosity is covered, so that the theory can be applied to the flying of birds, as well as to the swimming of fish or bacteria. As an example a system of three equal spheres moving along a common axis is studied.
\end{abstract}

\maketitle
\section{\label{1}Introduction}

The swimming of fish and the flying of birds continue to pose challenging theoretical problems. The physics of bird flight was first studied in detail by Otto Lilienthal in the nineteenth century \cite{1}. Since then, significant progress has been made in many years of dedicated research \cite{2}-\cite{5}.

The goal of theory is to calculate the time-averaged speed and power for given periodic shape variations of the body, at least for a simple model system. It is assumed that the motion of the fluid is well described by the Navier-Stokes equations for an incompressible viscous fluid. On average over a period the force exerted by the body on the fluid vanishes, so that thrust and drag cancel. In early work by Lighthill \cite{6} and Wu \cite{7} the thrust and power were calculated approximately as functions of the speed on the basis of potential flow theory for a planar strip. This work and subsequent developments have been reviewed by Childress \cite{3}, by Wu \cite{8},\cite{9}, and by Sparenberg \cite{10}. However, an independent calculation of the mean speed for given periodic shape variations is still lacking. Measurement of the power consumption has led to a surprisingly small friction coefficient, much smaller than that of an inert body, as was first observed by Gray \cite{11}.

It was first shown by Taylor \cite{12} that in the slow swimming of a microorganism the calculation of thrust can be circumvented. In this limiting case one can use the time-independent Stokes equations. The mean swimming velocity and mean rate of dissipation then follow from a purely kinematic calculation \cite{13},\cite{14}. For small amplitude swimming both quantities are quadratic in the amplitude of the stroke to lowest order. For a simple system, where the body is modeled as an assembly of rigid spheres held together by direct interaction forces and subject to periodic actuating forces which sum to zero, we have shown that in the high viscosity limit the swimming velocity and power can be calculated for any amplitude of stroke from kinematics alone \cite{15},\cite{16}.

In the following we investigate questions of thrust, velocity, and power for swimming or flying in a fluid of any viscosity, including the limit of low viscosity, for the same mechanical model as before. We assume for simplicity that the spheres experience Stokes friction. In addition we incorporate hydrodynamic interactions via virtual mass effects, as found from potential flow theory. We use Hamilton's equations of motion with added damping terms. In the limit of high viscosity, where resistive forces dominate, the earlier results are recovered. The model provides valuable insight also in the limit of low viscosity, where reactive forces dominate. In that regime the motion is dominated by virtual mass effects. Bernoulli forces and modified linear friction should be taken into account in a more realistic model. Nonetheless, the principle of the calculation, which exploits elimination of the fluid degrees of freedom, remains valid.

The flow is assumed to be laminar at all times. It is now realized that the boundary layer of swimming fish is laminar even at high Reynolds number \cite{9}. Virtual mass effects were discussed earlier by Lighthill \cite{17}. The numerical modeling of animal swimming and flight was reviewed by Deng et al. \cite{18}.

As an example a system of three equal spheres moving along a common axis is studied. For this simple system the mean swimming speed and mean power to second order in the amplitude of stroke can be evaluated analytically. The solution to a corresponding eigenvalue problem provides the optimal stroke to this order, as we found elsewhere in the resistive regime \cite{15}.

In our model the mean thrust, i.e. the frictional force exerted on the fluid averaged over a period in periodic swimming, vanishes identically. We find that the velocity of the geometric center of the assembly is driven by a different force, which we call the impetus. It has both a reactive and a resistive component. The impetus determines the center velocity with retardation. The mean impetus does not vanish.

It is known for small amplitude swimming in the resistive regime that the mean power is directly proportional to the mean velocity. We find for our example that the relation between mean power and mean velocity is nearly linear also for large amplitude swimming. Presumably the near linearity holds also for other systems in the whole regime of viscosity. If true, this would resolve the so-called Gray paradox \cite{9}, which is based on the mistaken notion that the power is quadratic in the velocity, as in Stokes friction.

\section{\label{2}Equations of motion}

We consider a set of $N$ rigid spheres of radii $a_1,...,a_N$ and masses $m_{p1},...,m_{pN}$, centered at positions $\du{R}=(\vc{R}_1,...,\vc{R}_N)$, and immersed in an incompressible viscous fluid of shear viscosity $\eta$ and mass density $\rho$. The fluid is of infinite extent in all directions. The flow velocity $\vc{v}(\vc{r},t)$ and pressure $p(\vc{r},t)$ of the fluid are assumed to satisfy the Navier-Stokes equations
\begin{equation}
\label{2.1}\rho\;\big[\frac{\partial\vc{v}}{\partial t}+\vc{v}\cdot\nabla\vc{v}\big]=\eta\nabla^2\vc{v}-\nabla p,\qquad\nabla\cdot\vc{v}=0.
\end{equation}
The flow velocity $\vc{v}$ is assumed to satisfy the no-slip boundary condition on the surface of the spheres.
The fluid is set in motion by time-dependent motions of the
spheres. At each time $t$ the velocity field $\vc{v}(\vc{r},t)$ tends to zero at infinity, and the pressure $p(\vc{r},t)$ tends to the constant ambient pressure $p_0$.

As the spheres move in the fluid they experience a frictional force. In addition there may be applied forces $\du{E}(t)=(\vc{E}_1(t),...,\vc{E}_N(t))$ and direct interaction forces which depend on the relative positions $\{\vc{R}_j-\vc{R}_k\}$ of sphere centers. We shall assume that the sum of applied forces vanishes, so that
\begin{equation}
\label{2.2}\sum^N_{j=1}\vc{E}_j(t)=0.
\end{equation}
The sum of direct interaction forces vanishes owing to Newton's third law. We assume that the frictional forces are linear in the sphere velocities, as given by low Reynolds number hydrodynamics on the slow timescale \cite{19}. The spheres are freely rotating so that there are no frictional torques.

The forces exerted by pressure gradients resist instantaneous acceleration and give rise to virtual mass effects \cite{20}. For a single sphere $j$ immersed in infinite fluid the virtual mass would be $\frac{1}{2}m_{fj}$, where $m_{fj}=4\pi\rho a_j^3/3$ is the mass of fluid displaced by the sphere. The virtual mass effect for a collection of $N$ spheres is embodied in a $(3N\times 3N)$-dimensional mass matrix $\du{m}$. This can be derived from potential flow theory by considering the irrotational flow pattern generated instantaneously by a set of sudden impulses $\du{S}=(\vc{S}_1,...,\vc{S}_N)$ from a state of rest. The total corresponding kinetic energy is
\begin{equation}
\label{2.3}\mathcal{K}=\frac{1}{2}\;\du{U}\cdot\du{m}\cdot\du{U},
\end{equation}
where $\du{U}=(\vc{U}_1,...,\vc{U}_N)$ is the set of sphere velocities. The kinetic energy is a sum of two contributions, $\mathcal{K}_p+\mathcal{K}_f$, with
\begin{equation}
\label{2.4}\mathcal{K}_p=\frac{1}{2}\sum^N_{j=1}m_{pj}U_j^2,\qquad\mathcal{K}_f=\frac{1}{2}\rho\int_{V_f}(\nabla\phi)^2\;d\vc{r},
\end{equation}
where the integration is over the part of space occupied by fluid and $\phi(\vc{r})$ is the velocity potential corresponding to the set of instantaneous velocities $\du{U}$. The potential $\phi(\vc{r})$ is linear in the sphere velocities $\du{U}=(\vc{U}_1,...,\vc{U}_N)$, so that the kinetic energy $\mathcal{K}_f$ is a quadratic form in $\du{U}$. The contribution $\mathcal{K}_f$ to the total kinetic energy defines the virtual mass. This depends parametrically on the positions $\du{R}$, leading to corresponding hydrodynamic interactions.

Besides the mass matrix $\du{m}$ it will be useful to define also the mobility matrix $\vc{\mu}$, the friction matrix $\vc{\zeta}$, and the inverse mass matrix $\du{w}$,
\begin{equation}
\label{2.5}\vc{\mu}=\vc{\zeta}^{-1},\qquad\du{w}=\du{m}^{-1}.
\end{equation}
 The four matrices are symmetric and depend on the relative positions of the sphere centers. The sphere momenta, including the virtual mass contribution, are given by
 \begin{equation}
\label{2.6}\du{p}=\du{m}\cdot\du{U},\qquad\du{U}=\du{w}\cdot\du{p}.
\end{equation}
Correspondingly we postulate the equations of motion
 \begin{equation}
\label{2.7}\frac{d\du{R}}{dt}=\du{U},\qquad\frac{d\du{p}}{dt}=-\frac{\partial\mathcal{K}}{\partial\du{R}}-\vc{\zeta}\cdot\du{U}-\frac{\partial V_{int}}{\partial\du{R}}+\du{E},
\end{equation}
where $\mathcal{K}$ is given by $\mathcal{K}=\frac{1}{2}\du{p}\cdot\du{w}\cdot\du{p}$ and $V_{int}$ is the potential of direct interaction forces. The partial derivative $\partial/\partial\du{R}$ is taken at constant momenta $\du{p}$. It is clear that the equations of motion (2.7) have only limited validity. The frictional forces are assumed to be linear in the sphere velocities, and Basset memory forces are neglected. Nonetheless it is of interest to explore the consequences of the equations as they stand.

We note that it follows from Eq. (2.7) that total dressed sphere momentum $\du{P}=\Sigma_j\du{p}_j$ is not conserved but changes due to friction with the fluid. If an impulse $\du{S}$ is imparted at time $t=0$ to the spheres in a state of rest due to applied forces $\du{E}=\du{S}\delta(t)$, then part of the momentum is transferred instantaneously to the fluid, reducing the sphere velocities to $\du{U}(0+)=\du{w}\cdot\du{S}$, as can be seen by integration over an infinitesimal time interval about $t=0$ and use of Eq. (2.6). The remaining momentum of the spheres is transferred to the fluid in the course of time by friction. The total momentum of spheres and fluid is conserved at all times.

\section{\label{3}Impetus, center velocity, and rate of dissipation}

We are looking for a solution of the equations of motion where the center of the assembly moves on average with uniform translational velocity and the individual spheres perform periodic motions about the moving center. The velocity $\vc{U}(t)$ of the geometric center is defined by
\begin{equation}
\label{3.1}U_\alpha=\frac{1}{N}\;\du{U}\cdot\du{u}_\alpha,\qquad\alpha=(x,y,z)
\end{equation}
where the symbol $\du{u}_x$ denotes a $3N$-dimensional vector with $1$ on the $x$ positions, $0$ on the $y,z$ positions, and cyclic. The $3N$-dimensional displacement vector $\du{d}(t)=(\vc{\delta}_1(t),...,\vc{\delta}_N(t))$ is defined by
\begin{equation}
\label{3.2}\vc{R}_j(t)=\vc{R}_{j0}+\int^t_0\vc{U}(t')\;dt'+\vc{\delta}_j(t),\qquad j=1,...,N,
\end{equation}
with the property
\begin{equation}
\label{3.3}\du{d}(t)\cdot\du{u}_\alpha=0,\qquad\alpha=(x,y,z),
\end{equation}
and with $\{\vc{R}_{j0}\}$ a set of equilibrium positions for which the direct interaction forces vanish.
Correspondingly the velocity vector is decomposed as
\begin{equation}
\label{3.4}\du{U}=U_\beta\du{u}_\beta+\dot{\du{d}}.
\end{equation}
Summation over repeated greek indices is implied. By substitution into Eqs. (2.6) and (2.7) we find
\begin{equation}
\label{3.5}\frac{d}{dt}[\du{m}\cdot(U_\beta\du{u}_\beta+\dot{\du{d}})]+\frac{\partial\mathcal{K}}{\partial\du{R}}+\vc{\zeta}\cdot(U_\beta\du{u}_\beta+\dot{\du{d}})+\frac{\partial V_{int}}{\partial\du{R}}=\du{E}(t).
\end{equation}
Multiplying from the left with the vector $\du{u}_\alpha$ we obtain
\begin{equation}
\label{3.6}\frac{d}{dt}\big(\mathcal{M}_{\alpha\beta}U_\beta\big)+\frac{d}{dt}\big(\du{u}_\alpha\cdot\du{m}\cdot\dot{\du{d}}\big)
+Z_{\alpha\beta}U_\beta+\du{u}_\alpha\cdot\vc{\zeta}\cdot\dot{\du{d}}=0,
\end{equation}
with mass tensor $\vc{\mathcal{M}}$ and friction tensor $\vc{Z}$ defined by
\begin{equation}
\label{3.7}\mathcal{M}_{\alpha\beta}=\du{u}_\alpha\cdot\du{m}\cdot\du{u}_\beta,\qquad Z_{\alpha\beta}=\du{u}_\alpha\cdot\vc{\zeta}\cdot\du{u}_\beta.
\end{equation}
We have used the fact that $\du{m}$ depends only on relative coordinates, so that $\du{u}_\alpha\cdot\partial\mathcal{K}/\partial{\du{R}}=0$.
Also we have used Newton's third law and Eq. (2.2). We note that in Eq. (3.6) the center velocity $U_\beta$ occurs only in the two places explicitly shown.

We rewrite Eq. (3.6) as
\begin{equation}
\label{3.8}\frac{d}{dt}\big(\vc{\mathcal{M}}\cdot\vc{U}\big)
+\vc{Z}\cdot\vc{U}=\vc{\mathcal{I}},
\end{equation}
with time-dependent impetus $\vc{\mathcal{I}}(t)$ given by
\begin{equation}
\label{3.9}\mathcal{I}_\alpha(t)=-\frac{d}{dt}\big(\du{u}_\alpha\cdot\du{m}\cdot\dot{\du{d}}\big)
-\du{u}_\alpha\cdot\vc{\zeta}\cdot\dot{\du{d}}.
\end{equation}
The mean impetus, averaged over a period $\tau$,
\begin{equation}
\label{3.10}\overline{\mathcal{I}_\alpha}=\frac{1}{\tau}\int^\tau_0\mathcal{I}_\alpha(t)\;dt=
-\du{u}_\alpha\cdot\overline{\vc{\zeta}\cdot\dot{\du{d}}},
\end{equation}
does not vanish, even though the total mean force exerted by the fluid on the assembly does vanish. The drag exerted on the assembly by the fluid is
\begin{equation}
\label{3.11}D_\alpha=-\du{u}_\alpha\cdot\vc{\zeta}\cdot\du{U}=-Z_{\alpha\beta}U_\beta-\du{u}_\alpha\cdot\vc{\zeta}\cdot\dot{\du{d}}.
\end{equation}
The thrust $\vc{T}$ is equal and opposite to the drag, $\vc{T}=-\vc{D}$. There are frictional forces on the spheres, but on time average the total drag vanishes, $\overline{\vc{D}}=\vc{0},\;\overline{\vc{T}}=\vc{0}$, as follows from Eqs. (2.2), (2.7) and Newton's third law for the interaction forces. We use the fact that the integral of $\du{u}_\alpha\cdot d\du{p}/dt$ over a period vanishes by periodicity, as well as the property $\du{u}_\alpha\cdot\partial\mathcal{K}/\partial{\du{R}}=0$, mentioned below Eq. (3.7). From $\overline{\vc{D}}=\vc{0}$ it follows that the mean impetus can also be expressed as
\begin{equation}
\label{3.12}\overline{\mathcal{I}_\alpha}=\overline{Z_{\alpha\beta}U_\beta}.
\end{equation}

We may regard Eq. (3.8) as a balance between central and internal motion. The equation must be solved for the center velocity $\vc{U}(t)$ for given impetus $\vc{\mathcal{I}}(t)$, which can be calculated from the displacement vector $\du{d}(t)$. In the resistive limit the total drag vanishes at any time \cite{21} and the reactive forces vanish, so that then $U_\alpha=M_{\alpha\beta}\mathcal{I}_\beta=-M_{\alpha\beta}\du{u}_\beta\cdot\vc{\zeta}\cdot\dot{\du{d}}$.

The displacement vector $\du{d}(t)$ may be calculated from displacements in relative space by using a transformation to center and relative coordinates. The geometric center of the assembly is given by
\begin{equation}
\label{3.13}\vc{R}=\frac{1}{N}\sum_{j=1}^N\vc{R}_j=\frac{1}{N}\;\vc{e}_\alpha\du{u}_\alpha\cdot\du{R}
\end{equation}
with Cartesian unit vectors $\vc{e}_\alpha$. We define relative coordinates $\{\vc{r}_j\}$ with $j=1,...,N-1$ as
  \begin{eqnarray}
\label{3.14}\vc{r}_1&=&\vc{R}_2-\vc{R}_1,\qquad\vc{r}_2=\vc{R}_3-\vc{R}_2,\qquad ...,\nonumber\\
\vc{r}_{N-1}&=&\vc{R}_N-\vc{R}_{N-1},
\end{eqnarray}
and denote the corresponding $(3N-3)$-vector $\du{r}=(\vc{r}_1,...,\vc{r}_{N-1})$. The $3N$-vector $(\vc{R},\du{r})$ is related to the vector $\du{R}$ by a transformation matrix $\du{T}$ according to
\begin{equation}
\label{3.15}(\vc{R},\du{r})=\du{T}\cdot\du{R}
\end{equation}
with explicit form given by Eqs. (3.13) and (3.14). The displacement vector $\du{d}$ is derived from the displacement $\vc{\xi}$ in relative space as
\begin{equation}
\label{3.16}\du{d}=\du{T}^{-1}\cdot(\vc{0},\vc{\xi}).
\end{equation}
Therefore the impetus is determined by displacements in relative space.

The time-dependent rate of dissipation can be expressed in the same matrix formalism. The rate of dissipation is given by
\begin{equation}
\label{3.17}\mathcal{D}=\du{U}\cdot\vc{\zeta}\cdot\du{U}.
\end{equation}
Once the center velocity $\vc{U}(t)$ has been calculated for known displacements we can also calculate the time-dependent rate of dissipation $\mathcal{D}(t)$ by use of Eq. (3.4). However, we derive an alternative expression which will be useful in the following. We can solve for the product $\vc{\zeta}\cdot\du{U}$ from the equation of motion
\begin{equation}
\label{3.18}\frac{d}{dt}\big(\du{m}\cdot\du{U}\big)+\vc{\zeta}\cdot\du{U}=\du{F},
\end{equation}
where $\du{F}$ is the vector of mechanical forces acting on the spheres,
\begin{equation}
\label{3.19}\du{F}=-\frac{\partial\mathcal{K}}{\partial\du{R}}-\frac{\partial V_{int}}{\partial\du{R}}+\du{E},
\end{equation}
which has the property $\du{u}_\alpha\cdot\du{F}=0$. Using this property we find for the rate of dissipation
\begin{equation}
\label{3.20}\mathcal{D}=\du{U}\cdot\du{F}-\du{U}\cdot\frac{d}{dt}\big(\du{m}\cdot\du{U}\big)=\dot{\du{d}}\cdot\du{F}-\du{U}\cdot\frac{d}{dt}\big(\du{m}\cdot\du{U}\big).
\end{equation}
Using here Eq. (3.18) again we can rewrite this as
\begin{equation}
\label{3.21}\mathcal{D}=\dot{\du{d}}\cdot\vc{\zeta}\cdot\dot{\du{d}}+U_\alpha\dot{\du{d}}\cdot\du{f}_\alpha-U_\alpha\du{u}_\alpha\cdot\frac{d}{dt}\big(\du{m}\cdot\du{U}\big),
\end{equation}
with friction vector $\du{f}_\alpha=\vc{\zeta}\cdot\du{u}_\alpha$. This generalizes an expression derived earlier in the resistive limit \cite{16}.

\section{\label{4}Small amplitude swimming}

For vanishing displacements Eq. (3.8) has the solution $\vc{U}=\vc{0}$. By formal series expansion in powers of the displacement vector $\du{d}$ we obtain a corresponding expansion of the center velocity
\begin{equation}
\label{4.1}\vc{U}=\vc{U}^{(1)}+\vc{U}^{(2)}+\vc{U}^{(3)}+....
\end{equation}
The first order velocity $\vc{U}^{(1)}$ satisfies the equation
\begin{equation}
\label{4.2}\mathcal{M}^0_{\alpha\beta}\frac{dU^{(1)}_\beta}{dt}+Z^0_{\alpha\beta}U^{(1)}_\beta=-\du{u}_\alpha\cdot\du{m}^0\cdot\ddot{\du{d}}
-\du{u}_\alpha\cdot\vc{\zeta}^0\cdot\dot{\du{d}},
\end{equation}
where the superscript $0$ indicates that the quantity is calculated for the configuration $\du{R}_0$. In particular for oscillating displacement, in complex notation
\begin{equation}
\label{4.3}\du{d}(t)=\mathrm{Re}\;[\du{d}_\omega e^{-i\omega t}],
\end{equation}
we find correspondingly
\begin{equation}
\label{4.4}U^{(1)}_{\alpha\omega}=\big[-i\omega\vc{\mathcal{M}}^0+\vc{Z}^0\big]^{-1}_{\alpha\beta}\du{u}_\beta\cdot(\omega^2\du{m}^0+i\omega\vc{\zeta}^0)\cdot\du{d}_\omega.
\end{equation}
In this situation the first order velocity oscillates in time and vanishes on time average.

Multiplying Eq. (3.8) by the mobility tensor $\vc{M}=\vc{Z}^{-1}$ and expanding to second order we obtain a more complicated equation for the second order velocity $\vc{U}^{(2)}$. It suffices to derive an expression for the time-average over a period $\tau=2\pi/\omega$,
\begin{equation}
\label{4.5}\overline{\vc{U}^{(2)}}=\frac{1}{\tau}\int^\tau_0\vc{U}^{(2)}(t)\;dt.
\end{equation}
Using periodicity and the first order equation Eq. (4.2) we obtain the expression
\begin{equation}
\label{4.6}\overline{U_\alpha^{(2)}}=-M^0_{\alpha\beta}\du{u}_\beta\cdot\overline{\vc{\zeta}^{(1)}\cdot\dot{\du{d}}}+\overline{M^{(1)}_{\alpha\beta}Z^0_{\beta\gamma}U^{(1)}_\gamma}.
\end{equation}
Alternatively the expression can be derived directly from Eq. (3.12) by use of $\vc{Z}^{(1)}=-\vc{Z}^0\vc{M}^{(1)}\vc{Z}^0$. Here we can use
\begin{equation}
\label{4.7}\vc{\zeta}^{(1)}=\du{d}\cdot\vc{\nabla}\vc{\zeta}\big|_0,\qquad\vc{M}^{(1)}=\du{d}\cdot\vc{\nabla}\vc{M}\big|_0,
\end{equation}
where $\vc{\nabla}$ indicates the gradient operator in configuration space, and the notation $\big|_0$ indicates that the value of the gradient is taken at $\du{R}_0$. The time average in the first term in Eq. (4.6) can be expressed as
\begin{equation}
\label{4.8}\du{u}_\beta\cdot\overline{\vc{\zeta}^{(1)}\cdot\dot{\du{d}}}=\frac{1}{2}\;\mathrm{Re}\;[-i\omega\du{d}^*_\omega\cdot\du{D}^\beta\big|_0\cdot\du{d}_\omega],
\end{equation}
with derivative friction matrix
\begin{equation}
\label{4.9}\du{D}^\beta=\vc{\nabla}\du{f}_\beta,\qquad\du{f}_\beta=\vc{\zeta}\cdot\du{u}_\beta,
\end{equation}
as introduced earlier \cite{16}.

In the second term in Eq. (4.6) we use the identity
\begin{equation}
\label{4.10}Z_{\alpha\gamma}M_{\gamma\beta}=\delta_{\alpha\beta}
\end{equation}
to show that
\begin{equation}
\label{4.11}\vc{\nabla}M_{\alpha\beta}=-M_{\alpha\gamma}\du{g}^\gamma_\delta M_{\delta\beta}
\end{equation}
with gradient vectors
\begin{equation}
\label{4.12}\du{g}^\beta_\gamma=\vc{\nabla}Z_{\beta\gamma}=\du{D}^\beta\cdot\du{u}_\gamma.
\end{equation}
The first order velocity $U^{(1)}_\gamma$ is eliminated by use of Eq. (4.4).
Then the second order mean swimming velocity can be expressed as
\begin{equation}
\label{4.13}\overline{U^{(2)}_\alpha}=\frac{1}{2}\;\mathrm{Re}\big[\;i\omega\;\du{d}^*_\omega\cdot \du{V}^\alpha(\omega)\big{|}_0\cdot\du{d}_\omega\big],
\end{equation}
with frequency-dependent matrix $\du{V}^\alpha(\omega)$ given by
\begin{equation}
\label{4.14}\du{V}^\alpha(\omega)=M_{\alpha\beta}\breve{\du{D}}^{\beta}(\omega),
\end{equation}
with reduced derivative friction matrix
\begin{equation}
\label{4.15}\breve{\du{D}}^{\beta}(\omega)=\du{D}^\beta-\du{g}^\beta_\gamma Y_{\gamma\delta}(\omega)\du{f}_\delta(\omega),
\end{equation}
with admittance tensor
\begin{equation}
\label{4.16}\vc{Y}(\omega)=\big[-i\omega\vc{\mathcal{M}}+\vc{Z}\big]^{-1},
\end{equation}
and impedance vector
\begin{equation}
\label{4.17}\du{f}_\delta(\omega)=(-i\omega\du{m}+\vc{\zeta})\cdot\du{u}_\delta.
\end{equation}

The matrix $\breve{\du{D}}^{\beta}(\omega)$ has the property
\begin{equation}
\label{4.18}\breve{\du{D}}^{\beta}(\omega)\cdot\du{u}_\alpha=0.
\end{equation}
Since $\vc{\zeta}$ depends only on relative coordinates $\du{u}_\alpha\cdot\vc{\nabla}\vc{\zeta}=\du{0}$, and hence
\begin{equation}
\label{4.19}\du{u}_\alpha\cdot\du{D}^{\beta}=\du{0},\qquad\du{u}_\alpha\cdot\du{g}^\beta_\gamma=0.
\end{equation}
As a consequence
\begin{equation}
\label{4.20}\du{u}_\alpha\cdot\du{V}^\beta(\omega)=\du{0},\qquad\du{V}^\alpha(\omega)\cdot\du{u}_\beta=\du{0}.
\end{equation}
These properties generalize those derived earlier at zero frequency \cite{16}.

To second order in the displacements the last term in Eq. (3.21) can be written as
\begin{equation}
\label{4.21}U_\alpha\du{u}_\alpha\cdot\frac{d}{dt}\big(\du{m}\cdot\du{U}\big)\approx\frac{1}{2}\frac{d}{dt}\big(U^{(1)}_\alpha\du{u}_\alpha\cdot\du{m}^0\cdot U^{(1)}_\beta\du{u}_\beta\big)+
U^{(1)}_\alpha\du{u}_\alpha\cdot\du{m}^0\cdot\ddot{\du{d}}.
\end{equation}
In the average over a period the first term on the right does not contribute. Hence for periodic motion the mean second order rate of dissipation can be expressed as
\begin{equation}
\label{4.22}\overline{\mathcal{D}^{(2)}}=\frac{1}{\tau}\int^\tau_0\bigg[\dot{\du{d}}\cdot\vc{\zeta}^0\cdot\dot{\du{d}}+U^{(1)}_\alpha\dot{\du{d}}\cdot\du{f}^0_\alpha-
U^{(1)}_\alpha\du{u}_\alpha\cdot\du{m}^0\cdot\ddot{\du{d}}\bigg]\;dt.
\end{equation}
Substituting Eq. (4.4) we find
\begin{equation}
\label{4.23}\overline{\mathcal{D}^{(2)}}=\frac{1}{2}\;\omega^2\;\mathrm{Re}\;[\du{d}^*_\omega\cdot\du{P}(\omega)\cdot\du{d}_\omega],
\end{equation}
with the complex matrix
\begin{equation}
\label{4.24}\du{P}(\omega)=\vc{\zeta}^0-Y^0_{\alpha\beta}(\omega)\du{f}^0_\alpha(\omega)\du{f}^0_\beta(\omega).
\end{equation}
The matrix is symmetric and has the properties
\begin{equation}
\label{4.25}\du{u}_\alpha\cdot\mathrm{Re}[\du{P}(\omega)]=\du{0},\qquad\mathrm{Re}[\du{P}(\omega)]\cdot\du{u}_\alpha=\du{0}.
\end{equation}
The properties Eq. (4.20) and (4.25) allow us to reduce the dimension of the matrix description by three by introducing center and relative coordinates.

\section{\label{5}Velocity matrices and power matrix}

The transformation given by Eqs. (3.13) and (3.14) can be used to reduce the calculation of the second order swimming velocity and rate of dissipation to one in relative space.
The matrices $\du{V}^\alpha(\omega)$ and $\du{P}(\omega)$ are transformed to
\begin{equation}
\label{5.1}\du{V}^\alpha_T(\omega)=\du{T}\cdot\du{V}^\alpha(\omega)\cdot\du{T}^{-1},\qquad\du{P}_T(\omega)=\du{T}\cdot\du{P}(\omega)\cdot\du{T}^{-1}.
\end{equation}
The first three rows of $\du{T}$ consist of $\du{u}_\alpha/N$ and the first three columns of $\du{T}^{-1}$ consist of $\du{u}_\alpha$. It follows from the properties Eq. (4.20) and (4.25) that the first three rows and columns of the transformed matrices $\du{V}^\alpha_T(\omega)$ and $\du{P}_T(\omega)$ vanish identically. Hence in this representation we can drop the center coordinates and truncate the matrices by erasing the first three rows and columns. We denote the truncated $(3N-3)\times(3N-3)$-matrices as $\hat{\du{V}}_T^\alpha(\omega)$ and $\hat{\du{P}}_T(\omega)$ and define displacements $\vc{\xi}_\omega$ in relative space by
\begin{equation}
\label{5.2}(\vc{0},\vc{\xi}_\omega)=\du{T}\cdot\du{d}_\omega.
\end{equation}
With this notation the mean second order swimming velocity and rate of dissipation are given by
\begin{eqnarray}
\label{5.3}\overline{U^{(2)}_\alpha}&=&\frac{1}{2}\;\mathrm{Re}\;i\omega\vc{\xi}^*_\omega\cdot\du{C}_T\cdot\hat{\du{V}}_T^\alpha(\omega)\cdot\vc{\xi}_\omega,\nonumber\\ \overline{\mathcal{D}^{(2)}}&=&\frac{1}{2}\;\omega^2\;\mathrm{Re}\;\vc{\xi}^*_\omega\cdot\du{C}_T\cdot\hat{\du{P}}_T(\omega)\cdot\vc{\xi}_\omega,
\end{eqnarray}
with the matrix
\begin{equation}
\label{5.4}\du{C}_T=[\widetilde{\du{T}^{-1}}\cdot\du{T}^{-1}]\;\vc{\hat{}}.
\end{equation}
This $(3N-3)\times(3N-3)$ dimensional matrix consists of numerical coefficients and is obtained from the corresponding $3N\times 3N$ matrix by truncation, as indicated by the final hat symbol.

We rewrite the expressions in Eq. (5.3) in a more convenient form with vectors and matrices consisting of dimensionless numbers.
We introduce the complex dimensionless vector
 \begin{equation}
\label{5.5}\vc{\xi}^c=\frac{1}{b}\;\vc{\xi}_\omega,
 \end{equation}
 where $b$ is a typical length scale, and define
  \begin{eqnarray}
\label{5.6}\du{B}^\alpha&=&b\big[\big(\du{C}_T\cdot i\hat{\du{V}}_T^\alpha(\omega)\big|_0\big)^{s\prime}
+i\big(\du{C}_T\cdot i\hat{\du{V}}_T^\alpha(\omega)\big|_0\big)^{a\prime\prime}\big],\nonumber\\
\du{A}&=&\frac{1}{b\eta}\;[\big(\du{C}_T\cdot\hat{\du{P}}_T(\omega)\big|_0\big)^{s\prime}+i\big(\du{C}_T\cdot\hat{\du{P}}_T(\omega)\big|_0\big)^{a\prime\prime}\big],
\end{eqnarray}
where the superscript $s$ indicates the symmetric part, the superscript $a$ the antisymmetric part, the single prime the real part, and the double prime the imaginary part.
With the scalar product
  \begin{equation}
\label{5.7}(\vc{\xi}^c|\vc{\eta}^c)=\sum^{N-1}_{j=1}\vc{\xi}_j^{c*}\cdot\vc{\eta}^c_j
 \end{equation}
 the mean swimming velocity and mean rate of dissipation can then be expressed as
  \begin{equation}
\label{5.8}\overline{U^{(2)}_\alpha}=\frac{1}{2}\omega
b(\vc{\xi}^c|\du{B}^{\alpha}|\vc{\xi}^c),\qquad\overline{\mathcal{D}^{(2)}}=\frac{1}{2}\eta\omega^2b^3(\vc{\xi}^c|\du{A}|\vc{\xi}^c).
 \end{equation}
 The matrices $\du{B}^{\alpha}$ and $\du{A}$ are hermitian. We call $\du{B}^{\alpha}$ the velocity matrix and $\du{A}$ the power matrix.

 We ask for the stroke with maximum swimming velocity in a class of strokes with equal rate of dissipation for fixed values of the geometric parameters, fixed frequency $\omega$, and given values of viscosity $\eta$ and mass density $\rho$. Maximizing the quadratic form $\omega b^2\overline{U^{(2)}_\alpha}-\lambda^\alpha\overline{\mathcal{D}^{(2)}}$ with Lagrange multiplier $\lambda^\alpha$ we obtain the generalized eigenvalue problem
 \begin{equation}
\label{5.9}\du{B}^\alpha\vc{\xi}^c=\lambda^\alpha\du{A}\vc{\xi}^c.
 \end{equation}
 The eigenvalues $\{\lambda^\alpha\}$ are real. The maximum efficiency $E_T=\omega b^2|\overline{U^{(2)}_\alpha}|/\overline{\mathcal{D}^{(2)}}|$ for motion in direction $\alpha$ is given by the maximum eigenvalue as
   \begin{equation}
\label{5.10}E^\alpha_{Tmax}=\lambda^\alpha_{max}.
 \end{equation}
 The set $\{E^x_{Tmax},E^y_{Tmax},E^z_{Tmax}\}$ depends on the choice of Cartesian coordinate system. Further optimization may be possible by a rotation of axes. In particular cases a natural choice of axes will suggest itself.

 \section{\label{6}Power and dissipation}

 We view the swimmer or flyer as a dynamical system in periodic motion, driven by actuating forces $\du{E}(t)$ satisfying $\du{u}_\alpha\cdot\du{E}=0$ and $\du{E}(t+\tau)=\du{E}(t)$. In an expansion in powers of the actuating forces the first order displacements are given by Eq. (4.3). To second order the mean swimming velocity and mean rate of dissipation are given by Eq. (5.8). In this section we relate the mean power supplied by the actuating forces to the mean rate of dissipation.

The instantaneous power supplied by the actuating forces is given by
  \begin{equation}
\label{6.1}P(t)=\du{E}(t)\cdot\du{U}(t).
 \end{equation}
 From Eq. (2.7)
   \begin{equation}
\label{6.2}\frac{d}{dt}\big(\mathcal{K}+V_{int}\big)=
-\du{U}\cdot\vc{\zeta}\cdot\du{U}+\du{E}\cdot\du{U}.
 \end{equation}
 Hence we find for periodic motion
   \begin{equation}
\label{6.3}\overline{\du{U}\cdot\vc{\zeta}\cdot\du{U}}=\overline{\du{E}\cdot\du{U}}.
 \end{equation}
 This shows that on time average over a period the power is fully dissipated by friction.

 Since the mean thrust vanishes the usual definition of energy wastage \cite{2} makes no sense here. Instead we define the energy wastage $\mathcal{E}(t)$ as the difference
   \begin{equation}
\label{6.4}\mathcal{E}=\overline{P}-\overline{\vc{U}}\cdot\overline{\vc{\mathcal{I}}}.
 \end{equation}
 In the theory of fish swimming and bird flight the energy wastage has been associated with energy being lost to the vortical wake \cite{9}. We define the corresponding Froude efficiency as
 \begin{equation}
\label{6.5}\eta_F=\frac{\overline{\vc{U}}\cdot\overline{\vc{\mathcal{I}}}}{\overline{P}}.
 \end{equation}
 This concept may be useful for the comparison of different strokes.

\section{\label{7}Hydrodynamic interactions}

In order to perform explicit calculations we must specify the form of the hydrodynamic interactions appearing in the friction matrix and the mass matrix. In practice one uses approximate expressions which are presumed to be reasonably accurate in the range of distances considered.

The friction matrix can be calculated from an approximation to the mobility matrix based on Oseen's pair interaction \cite{19}. In this approximation the pair mobility tensor for the pair $(j,k)$ is given by
\begin{equation}
\label{7.1}\vc{\mu}_{jk}=\frac{1}{6\pi\eta a_j}\vc{1}\delta_{jk}
+\frac{1}{8\pi\eta}\bigg[\frac{\vc{1}}{|R_j-R_k|}+\frac{(\vc{R}_j-\vc{R}_k)(\vc{R}_j-\vc{R}_k))}{|R_j-R_k|^3}\bigg](1-\delta_{jk}).
\end{equation}
The mobility matrix $\vc{\mu}$ is composed of pair tensors, and the friction matrix $\vc{\zeta}$ is its inverse.

The calculation of the mass matrix $\du{m}$ is based on potential flow theory. A dipole approximation to the mass matrix can be evaluated on the basis of an expression for the force on a sphere in a uniform flow in potential flow theory, as given by Landau and Lifshitz \cite{22} and by Batchelor \cite{23}.

In potential flow theory the flow velocity is expressed as $\vc{v}=-\nabla\phi$ with a scalar potential $\phi$, which satisfies Laplace's equation $\nabla^2\phi=0$ by incompressibility. A sphere of radius $a$, centered at the origin and moving with velocity $\vc{U}$ in a fluid at rest generates a potential
 \begin{equation}
\label{7.2}\phi_U(\vc{r})=\frac{1}{2}\;a^3\frac{\vc{r}}{r^3}\cdot\vc{U},\qquad r>a,
\end{equation}
corresponding to the dipole moment
 \begin{equation}
\label{7.3}\vc{q}_U=\frac{1}{2}\;a^3\vc{U}.
\end{equation}
If the sphere is placed in a uniform flow $\vc{v}_0$ this is modified to \cite{24}
 \begin{equation}
\label{7.4}\vc{q}=\frac{1}{2}\;a^3(\vc{U}-\vc{v}_0).
\end{equation}
For a collection of $N$ spheres in a fluid at rest at infinity the velocities and dipole moments are related in dipole approximation as
 \begin{equation}
\label{7.5}\vc{U}_j=\frac{2}{a_j^3}\;\vc{q}_j+\sum_{k\neq j}\vc{F}_{jk}\cdot\vc{q}_k,\qquad j=1,...,N,
\end{equation}
with dipole interaction tensor
 \begin{equation}
\label{7.6}\vc{F}_{jk}=\vc{F}(\vc{R}_j-\vc{R}_k),\qquad\vc{F}(\vc{r})=\frac{-\vc{1}+3\hat{\vc{r}}\hat{\vc{r}}}{r^3},
\end{equation}
where $\hat{\vc{r}}=\vc{r}/r$. We abbreviate Eq. (7.5) as
 \begin{equation}
\label{7.7}\du{U}=\vc{\mathcal{A}}^{-1}\cdot\du{q},\qquad\du{q}=\vc{\mathcal{A}}\cdot\du{U}.
\end{equation}

The velocity of sphere $j$ after a sudden impulse $\vc{S}_j$ from a state of rest is given by \cite{22},\cite{23}
 \begin{equation}
\label{7.8}\big(m_{pj}+\frac{1}{2}m_{fj}\big)\vc{U}_j=\vc{S}_j+\frac{3}{2}m_{fj}\sum_{k\neq j}\vc{F}_{jk}\cdot\vc{q}_k,\qquad j=1,...,N.
\end{equation}
Substituting from Eq. (7.7) and solving for the velocities we find by use of Eq. (7.5) and $m_{fj}=4\pi\rho a_j^3/3$
 \begin{equation}
\label{7.9}\du{U}=\du{w}\cdot\du{S}
\end{equation}
with matrix
 \begin{equation}
\label{7.10}\du{w}=\big[\du{m}_p-\du{m}_f+4\pi\rho\vc{\mathcal{A}}\big]^{-1},
\end{equation}
where the matrices $\du{m}_{p}$ and $\du{m}_{f}$ are diagonal with elements $m_{pj}\vc{1}$ and $m_{fj}\vc{1}$.
The approximate effective mass matrix is
\begin{equation}
\label{7.11}\du{m}=\du{m}_p-\du{m}_f+4\pi\rho\vc{\mathcal{A}}\big .
\end{equation}
If the spheres are neutrally buoyant one has simply
 \begin{equation}
\label{7.12}\du{m}=4\pi\rho\vc{\mathcal{A}},\qquad\du{w}=\frac{1}{4\pi\rho}\vc{\mathcal{A}}^{-1}.
\end{equation}
In our application we shall consider this case.

\section{\label{8}Three-sphere swimmer or flyer}

 The simplest application of the theory is to a three-sphere swimmer or flyer with the three spheres aligned on the $x$ axis, as studied by Golestanian and Ajdari \cite{25} in the resistive limit. The spheres move along the $x$ axis, and the $y$ and $z$ coordinates can be ignored \cite{26}. There are only two relative coordinates $r_1=x_2-x_1$ and $r_2=x_3-x_2$, and the relevant parts of the matrices $\du{B}^x$ and $\du{A}$ are two-dimensional. The relevant parts are denoted as $\du{B}^{xx}$ and $\du{A}^x$. In the bilinear theory we consider a point $\du{r}_0$ in $\du{r}$-space with coordinates $(d_1,d_2)$, corresponding to the configuration $\du{R}_0$ of the rest system. As an example we consider the case of equal-sized spheres with $a_1=a_2=a_3=a$, equal masses $m_1=m_2=m_3=m=4\pi\rho a^3/3$, and equal distances between centers $d_1=d_2=d$.

For this case analytic expressions for the matrices $\du{B}^{xx}$ and $\du{A}^x$ can be derived, but are too complicated to be presented. In the high viscosity limit the expressions reduce to those derived previously \cite{15}. The matrices, defined with $b=a$ in Eq. (5.6), depend only on the ratio $d/a$ and the dimensionless viscosity \cite{27}
 \begin{equation}
\label{8.1}\eta_*=\frac{\eta}{\omega a^2\rho}.
\end{equation}
The two eigenvalues $\lambda_\pm=\pm\lambda_+$, as well as the corresponding eigenvectors $\vc{\xi}_\pm=(1,\xi_\pm)$ with $\xi_-=\xi^*_+$, also depend only on these two variables. The dependence on the viscosity $\eta_*$ is surprisingly slight over the whole range of $\eta_*$ values. The absolute value $|\xi_+|$ is close to unity over the whole range. In Fig. 1 we show the variation of $\lambda_+$ and $|\xi_+|$ with $\eta_*$ for $d=5a$. The argument of $\xi_+$ increases slightly from $0.6278\;\pi$ at $\eta_*=0$ to $0.6285\;\pi$ at $\eta_*=10^6$.

In the bilinear theory the optimal orbit $(r_1(t),r_2(t))$ in relative space is given by $\vc{r}(t)=\vc{r}_0+\vc{\xi}_0(t)$ with $\vc{r}_0=(d,d)$ and
\begin{equation}
\label{8.2}\vc{\xi}_0(t)=\varepsilon a\;\mathrm{Re}\;[\vc{\xi}_+\exp(-i\omega t)],
\end{equation}
with amplitude factor $\varepsilon$. The corresponding displacement vector in configuration space is given by
\begin{equation}
\label{8.3}\du{d}_0(t)=\du{T}^{-1}\cdot\left(\begin{array}{c}0\\\vc{\xi}_0(t)\end{array}\right),
\qquad\du{T}=\left(\begin{array}{ccc}\frac{1}{3}&\frac{1}{3}&\frac{1}{3}\\-1&1&0\\0&-1&1
\end{array}\right).
\end{equation}
In Fig. 2 we show a snapshot of the spheres and their velocities in the instantaneous rest frame at $t=0$ for $\varepsilon=3,\;d=5a$ and $\eta_*=0.01$.
In Fig. 1 of Ref. 15 we showed the elliptical orbit in relative space for $d=5a$ and $\varepsilon=0.1$ in the limit of high viscosity. In Fig. 3 we compare this with the corresponding orbit in the limit of zero viscosity, corresponding to small $\eta_*$. The two plots are indistinguishable on the scale of the figure.

From the periodic displacement $\du{d}_0(t)$ we can calculate the instantaneous swimming velocity $U(t)$ as a series of harmonics from Eq. (3.8). The zeroth harmonic yields the mean swimming velocity $\overline{U}$. From $U(t)$ and $\du{d}_0(t)$ we can calculate the time-dependent rate of dissipation $\mathcal{D}(t)$, and hence the mean $\overline{\mathcal{D}}$, by use of Eq. (3.17).

In Fig. 4 we show the reduced mean swimming velocity $\overline{U}/(\varepsilon^2\omega a)$ as a function of $\varepsilon$ for $d=5a$ and $\eta_*=0.01$. In Fig. 5 we show the reduced mean power $\overline{P}/(\varepsilon^2\eta\omega^2a^3)$ vs. the reduced mean swimming velocity $\overline{U}/(\varepsilon^2\omega a)$ in the range $0<\varepsilon<3$.  In Fig. 6 we show the efficiency $E_T=\eta\omega a^2\overline{U}/\overline{P}$ as a function of $\varepsilon$. The efficiency increases monotonically with the amplitude factor. In Fig. 7 we show the time-dependence of the impetus $\mathcal{I}(t)$ and the center velocity $U(t)$ as functions of time during a period for $\varepsilon=3$. We also plot separately the resistive contribution to the impetus. At this value of $\eta_*$ this is much smaller than the reactive part. It is noteworthy that the variations in time of the center velocity are much smaller than those of the impetus. The center velocity follows the impetus with an after-effect. In Fig. 8 we show the absolute value of the Fourier coefficients $f_n$ of the velocity $U(t)$, normalized to $f_0=1$, for $\varepsilon=3$. This shows that only a small number of harmonics contribute appreciably.

The equality of mean power and mean rate of dissipation given by Eq. (6.3) can be checked numerically. The Froude efficiency $\eta_F$, defined in Eq. (6.5), for these three values of $\varepsilon$ is $0.0004,\;0.0020,\;0.0057$, respectively.

It is of interest to compare the above results with values obtained from the numerical solution of the equations of motion Eq. (2.7) with hydrodynamic interactions given by Eqs. (7.1) and (7.12) and with prescribed oscillating actuating forces. In order to stabilize the system we consider microswimmers with internal harmonic interactions  \cite{28}.
In matrix form the forces may be expressed as
 \begin{equation}
\label{8.4}\du{F}=\du{H}\cdot(\du{R}-\du{R}_0)+\du{E},
 \end{equation}
 with a real symmetric matrix $\du{H}$ with the property $\du{H}\cdot\du{u}_\alpha=0$. The actuating forces $\{\vc{E}_j(t)\}$ can be chosen to correspond to the eigenvector with maximum eigenvalue in the problem Eq. (5.9). The first term in Eq. (8.4) represents a harmonic approximation to the direct interactions. We use harmonic interactions given by the $3\times 3$-matrix
  \begin{equation}
\label{8.5}\du{H}=k\left(\begin{array}{ccc}-1&1&0\\1&-2&1\\0&1&-1
\end{array}\right)
\end{equation}
with elastic constant $k$. This corresponds to nearest neighbor interactions of equal strength $k$ between the three spheres. The stiffness of the assembly is characterized by the dimensionless number $\sigma$ defined by
\begin{equation}
\label{8.6}\sigma=\frac{k}{\pi\eta a\omega}.
\end{equation}

We use the first order equations of motion,
 \begin{equation}
\label{8.7}\frac{d\du{R}^{(1)}}{dt}=\du{U}^{(1)},\qquad\du{m}^0\cdot\frac{d\du{U}^{(1)}}{dt}=-\vc{\zeta}^0\cdot\du{U}^{(1)}+\du{H}\cdot\du{R}^{(1)}+\du{E}_0,
\end{equation}
to calculate the actuating forces $\du{E}_0(t)$ corresponding to the optimal linear motion given by Eqs. (8.3) and (4.4).
These have the property $\du{u}_\alpha\cdot\du{E}_0=0$, so that the sum of actuating forces vanishes. We choose initial conditions corresponding to the rest configuration
 \begin{eqnarray}
\label{8.8}x_1(0)&=&0,\qquad x_2(0)=d,\qquad x_3(0)=2d,\nonumber\\
U_1(0)&=&0,\qquad U_2(0)=0,\qquad U_3(0)=0.
\end{eqnarray}

The kinetic energy term in Eq. (2.7) makes the direct numerical solution of the equations of motion time-consuming. Instead we use an iterative procedure, neglecting the kinetic energy term in first approximation. Then the equations can be solved by a fast procedure. From the solution the kinetic forces, defined as $-\partial\mathcal{K}/\partial\du{R}$, can be calculated as a function of time. In the next step we include the kinetic forces and repeat the procedure. The kinetic forces are small compared to the actuating forces and the solution converges after a few iterations.

In Fig. 9 we show the numerical solution of the equations of motion Eqs. (2.6) and (2.7) with forces given by Eq. (8.4) for $d=5a$, viscosity $\eta_*=0.01$, stiffness $\sigma=10$, and amplitude factor $\varepsilon=1.5$ for the first fifty periods. In Fig. 10 we show the mean value of the kinetic energy for successive periods. The orbit for the last period hardly differs from the ellipse given by Eq. (8.3), as shown in Fig. 11. The mean swimming velocity and the mean rate of dissipation can be calculated as time-averages over the last period. The efficiency is $E_T=94\times 10^{-5}$, equal to the value calculated from the periodic orbit by use of Eq. (3.8) for displacements $\du{d}_0(t)$ with $\varepsilon=1.5$, shown in Fig. 6.

\section{\label{9}Discussion}

In general the performance of a swimmer or flyer can be measured in terms of the dimensionless efficiency $E_T$, defined as the ratio
\begin{equation}
\label{9.1}E_T=\frac{\eta\omega a^2\overline{U}}{\overline{P}},
\end{equation}
where $a$ is a conveniently chosen length scale, $\overline{U}$ is the mean speed and $\overline{P}$ is the mean power, averaged over a period $\tau=2\pi/\omega$.  We note that the lower bound of the efficiency $E_T$ vanishes, since for a periodic motion in relative phase space which is time-reversible the mean swimming velocity vanishes.
A striking result of the present analysis is that for the simple three-sphere model, for which analytic calculations can be performed, the maximum efficiency $E_{Tmax}$ is nearly independent of the dimensionless viscosity $\eta_*=\eta/(\omega a^2\rho)$, as shown in Fig. 1, see Eq. (5.10). As a consequence the mean speed is nearly inversely proportional to the shear viscosity $\eta$ for given power. We expect that this is a general feature of the mechanical system defined by equations of motion of the type Eq. (2.7) with hydrodynamic interactions as detailed in Sec. VII. This explains the great advantage that birds in air have over fish in water.

A second result of the analysis is that the mean power is equal to the mean rate of dissipation, as shown in Sec. VI. There is no additional energy loss related to the rate of change of the virtual mass.

The mean speed and mean power can be evaluated for more sophisticated model swimmers or flyers by similar analysis. Elsewhere we studied longer chains with both longitudinal and transverse excitation in the resistive limit \cite{26}. That analysis can be extended to the full range of viscosity, based on the equations of motion Eqs. (2.6) and (2.7).

For assumed periodic displacements the velocity of the assembly can be derived from the equation of motion Eq. (3.8) by decomposition in harmonics, as demonstrated in the three-sphere model. The mean power is found as a collateral of the calculation. The full range of viscosity is covered, so that the analysis provides an interesting link between the flying of birds, the swimming of fish, and the swimming of bacteria.

\newpage

\section*{Figure captions}

\subsection*{Fig. 1}
Plot of the eigenvalue $\lambda_+$ and the absolute value $|\xi_+|$ for the corresponding eigenvector $\vc{\xi}_+=(1,\xi_+)$ of the three-sphere model with $d=5a$ as functions of the dimensionless viscosity $\eta_*$.

\subsection*{Fig. 2}
Snapshot of the spheres and their velocities in the rest frame at $t=0$ for the motion given by Eq. (8.2) with $\varepsilon=3,\;d=5a$ and $\eta_*=0.01$.

\subsection*{Fig. 3}
Plot of the elliptical orbit in the $r_1r_2$ plane corresponding to Eq. (8.2) with $\varepsilon=0.1,\;d=5a$ and $\eta_*=0.01$ (solid curve).  We also plot the elliptical orbit for the high viscosity limit (dashed curve). The two curves cannot be distinguished on the scale of the figure.

\subsection*{Fig. 4}
Plot of the reduced mean swimming velocity $\overline{U}/(\varepsilon^2\omega a)$ for $d=5a$ and $\eta_*=0.01$ as a function of the amplitude $\varepsilon$ as calculated from Eq. (3.8) for displacements given by Eq. (8.3).

\subsection*{Fig. 5}
 Parametric plot of the reduced mean swimming power $\overline{\mathcal{P}}/(\varepsilon^2\eta\omega^2 a^3)$ vs. the reduced mean swimming velocity $\overline{U}/(\varepsilon^2\omega a)$ for $d=5a$, $\eta_*=0.01$ and $0<\varepsilon\leq 3$.

\subsection*{Fig. 6}
 As in Fig. 4 for the efficiency $E_T=\eta\omega a^2\overline{U}/\overline{\mathcal{D}}$.

 \subsection*{Fig. 7}

Plot of the impetus $\mathcal{I}(t)$ (solid curve), the resistive contribution to the impetus (short dashes) and of the center velocity $U(t)$ (long dashes) as functions of time during a period for the three-sphere swimmer for $d=5a$, $\eta_*=0.01$ and amplitude factor $\varepsilon=3$.

\subsection*{Fig. 8}

Plot of the absolute values of the Fourier coefficients $f_n$ of harmonics of frequency $n\omega$, normalized to $f_0=1$, of the center velocity $U(t)$ of the three-sphere swimmer for $d=5a$ and $\eta_*=0.01$ with amplitude factor $\varepsilon=3$.

 \subsection*{Fig. 9}

 Plot of the positions of the three spheres found by numerical integration of the equations of motion Eq. (2.7) for $d=5a$, $\eta_*=0.01,\;\sigma=10,\;\varepsilon=1.5$ for fifty periods of time.

 \subsection*{Fig. 10}

Plot of the mean value of the kinetic energy for successive periods $k=1,...,50$ corresponding to Fig. 9.

  \subsection*{Fig. 11}

Plot of the orbit during the last period of Fig.9 (solid curve) compared with the elliptical orbit given by Eq. (8.2) (dashed curve).

\newpage
\setlength{\unitlength}{1cm}
\begin{figure}
 \includegraphics{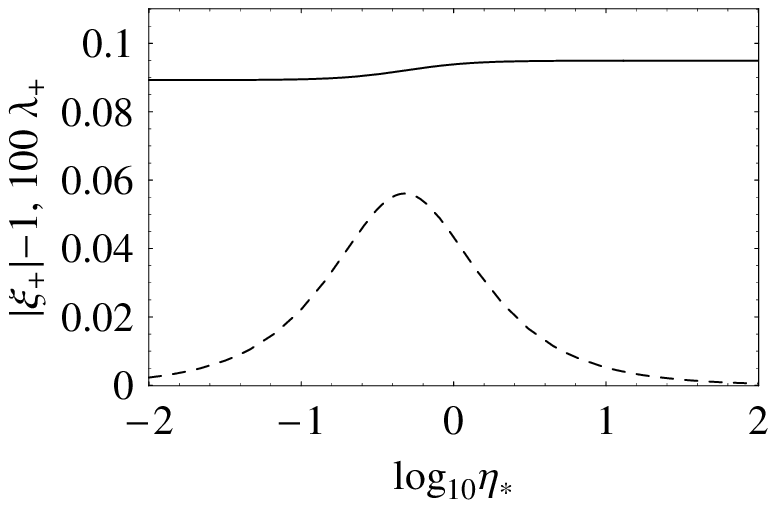}
   \put(-9.1,3.1){}
\put(-1.2,-.2){}
  \caption{}
\end{figure}
\newpage
\clearpage
\newpage
\setlength{\unitlength}{1cm}
\begin{figure}
 \includegraphics{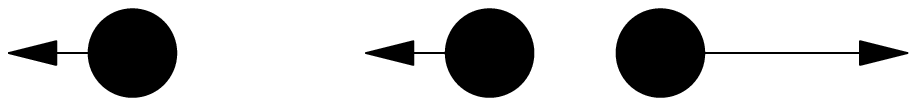}
   \put(-9.1,3.1){}
\put(-1.2,-.2){}
  \caption{}
\end{figure}
\newpage
\clearpage
\newpage
\setlength{\unitlength}{1cm}
\begin{figure}
 \includegraphics{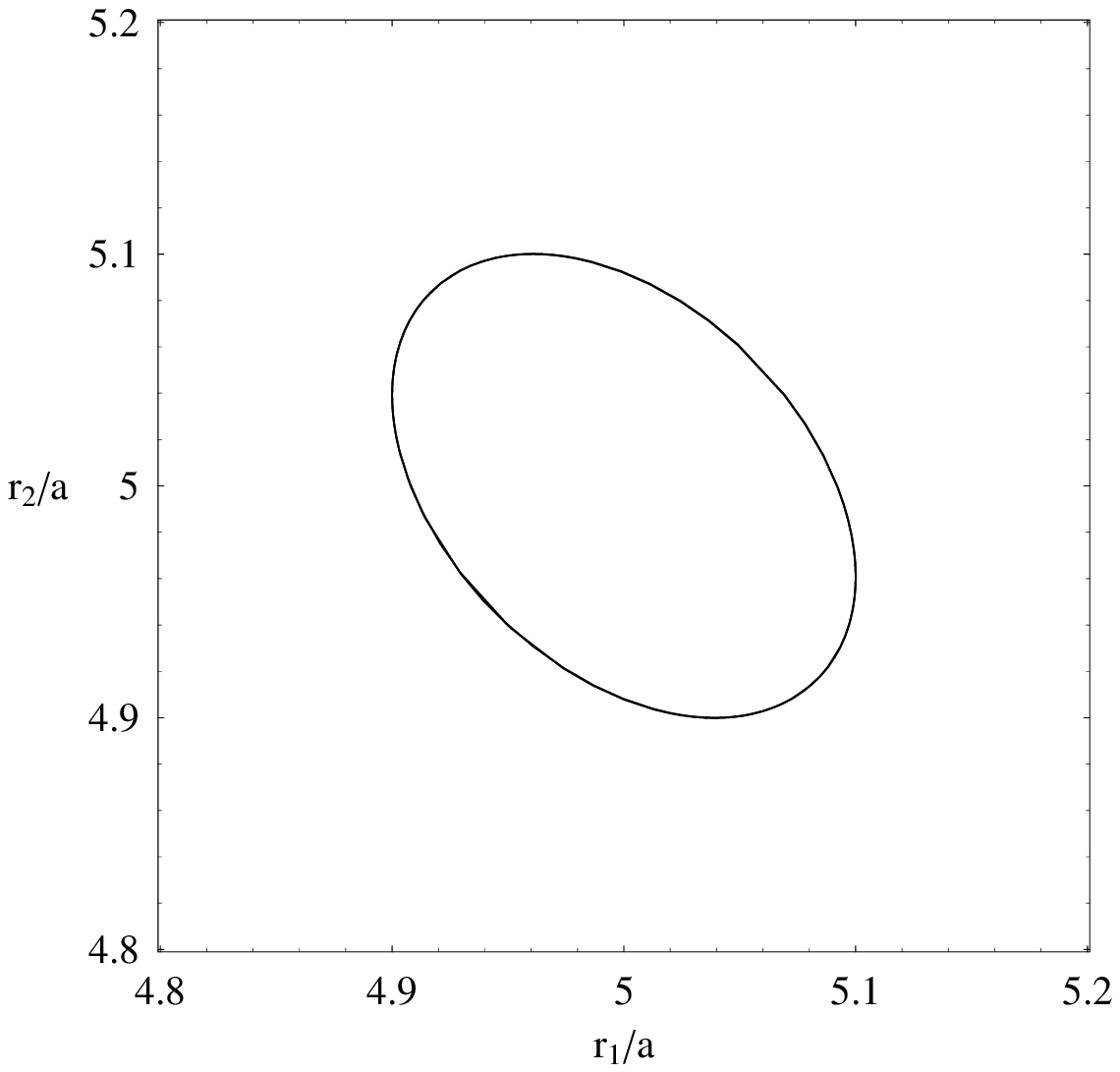}
   \put(-9.1,3.1){}
\put(-1.2,-.2){}
  \caption{}
\end{figure}
\newpage
\clearpage
\newpage
\setlength{\unitlength}{1cm}
\begin{figure}
 \includegraphics{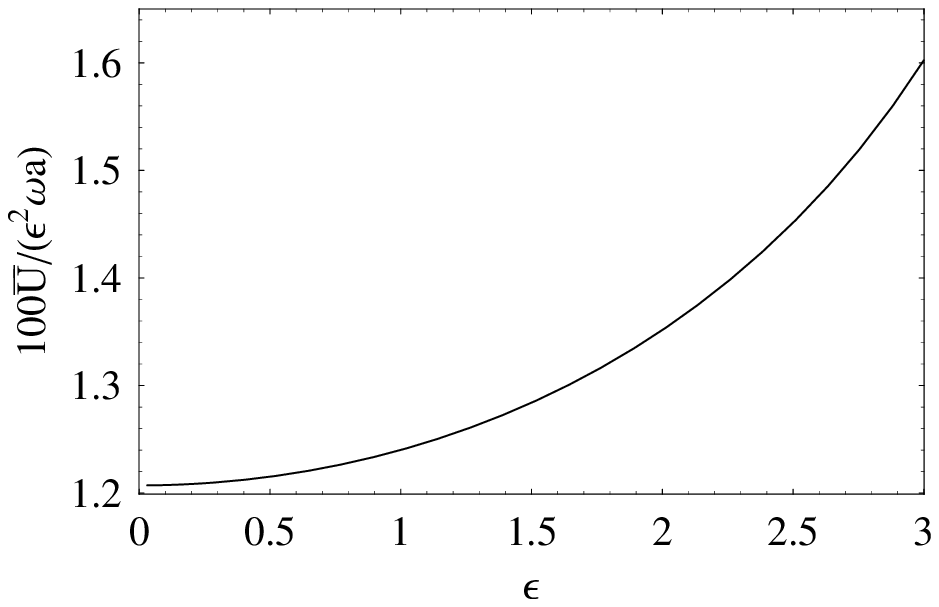}
   \put(-9.1,3.1){}
\put(-1.2,-.2){}
  \caption{}
\end{figure}
\newpage
\clearpage
\newpage
\setlength{\unitlength}{1cm}
\begin{figure}
 \includegraphics{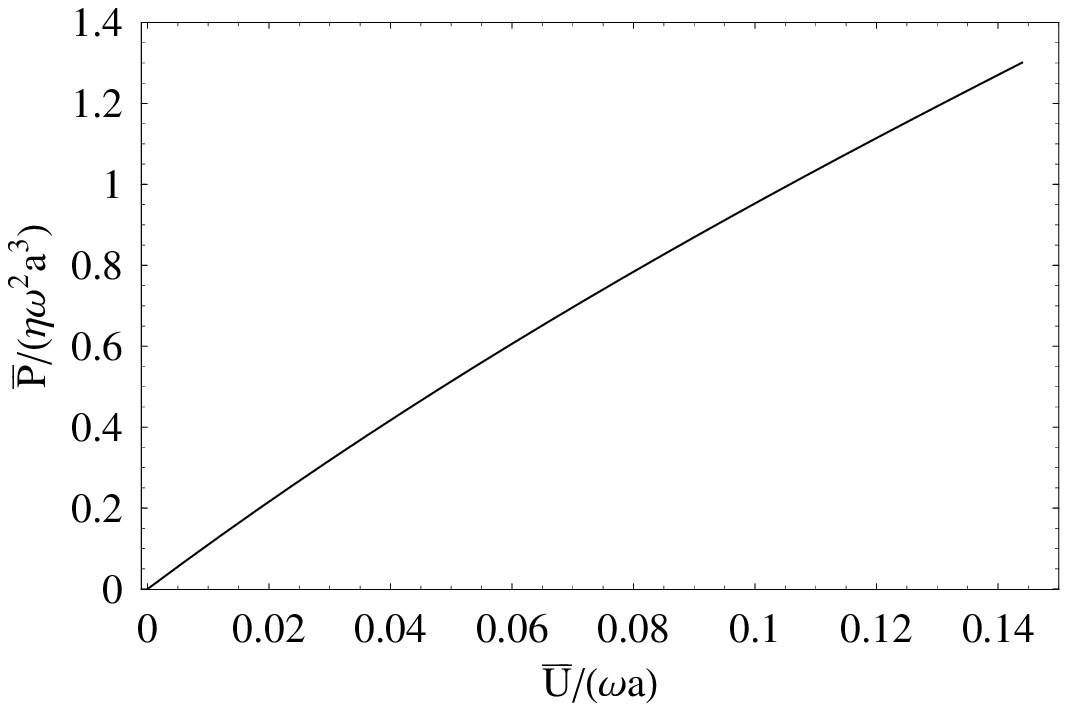}
   \put(-9.1,3.1){}
\put(-1.2,-.2){}
  \caption{}
\end{figure}
\newpage
\clearpage
\newpage
\setlength{\unitlength}{1cm}
\begin{figure}
 \includegraphics{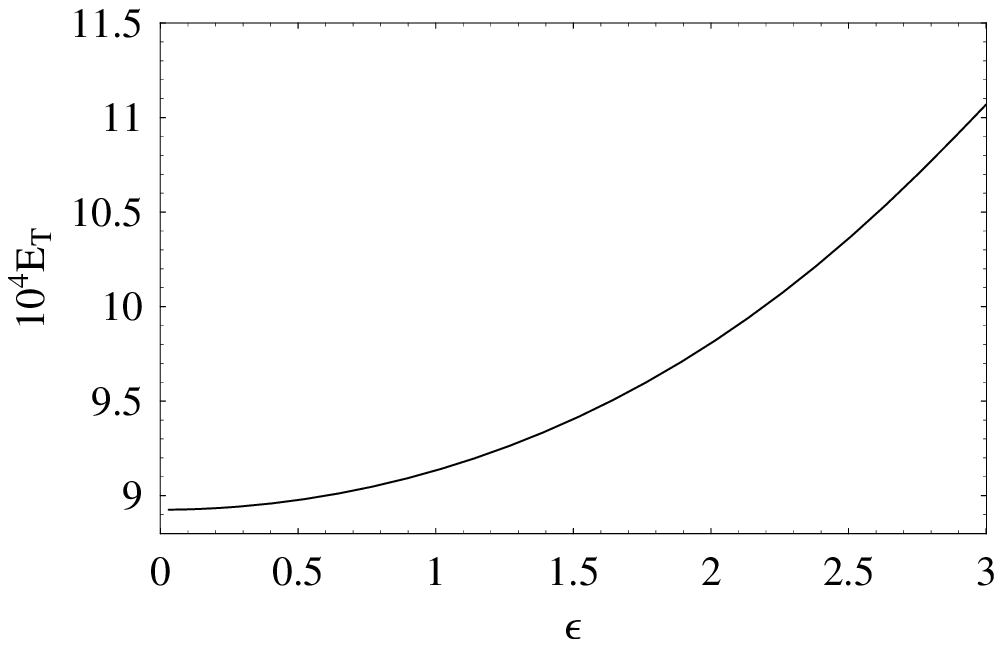}
   \put(-9.1,3.1){}
\put(-1.2,-.2){}
  \caption{}
\end{figure}
\newpage
\clearpage
\newpage
\setlength{\unitlength}{1cm}
\begin{figure}
 \includegraphics{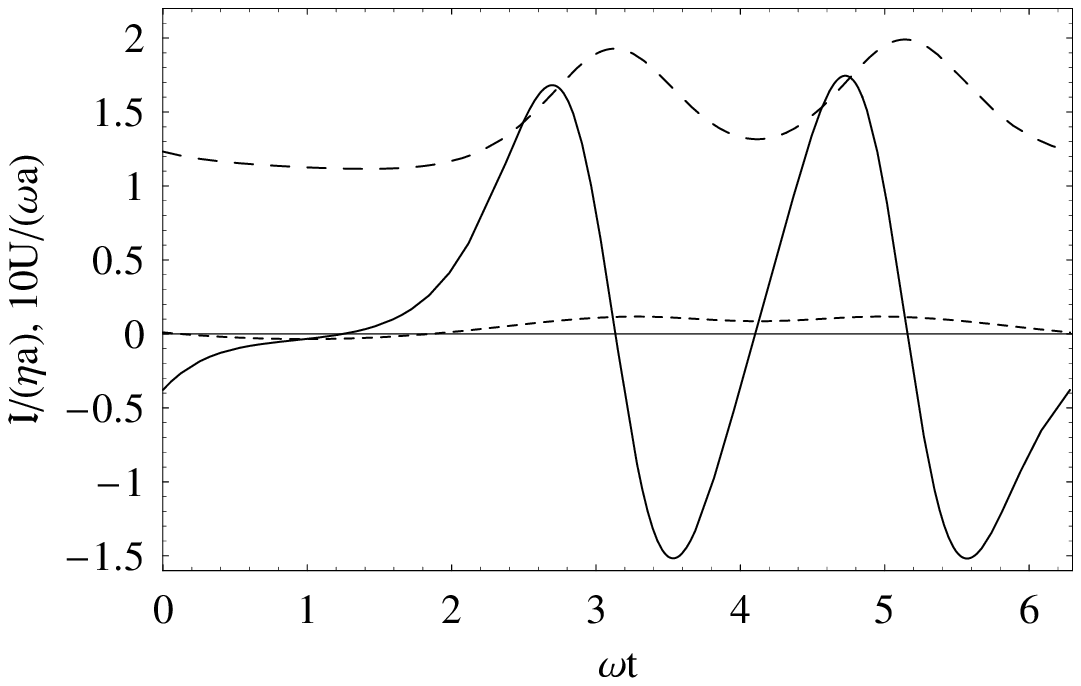}
   \put(-9.1,3.1){}
\put(-1.2,-.2){}
  \caption{}
\end{figure}
\newpage
\clearpage
\newpage
\setlength{\unitlength}{1cm}
\begin{figure}
 \includegraphics{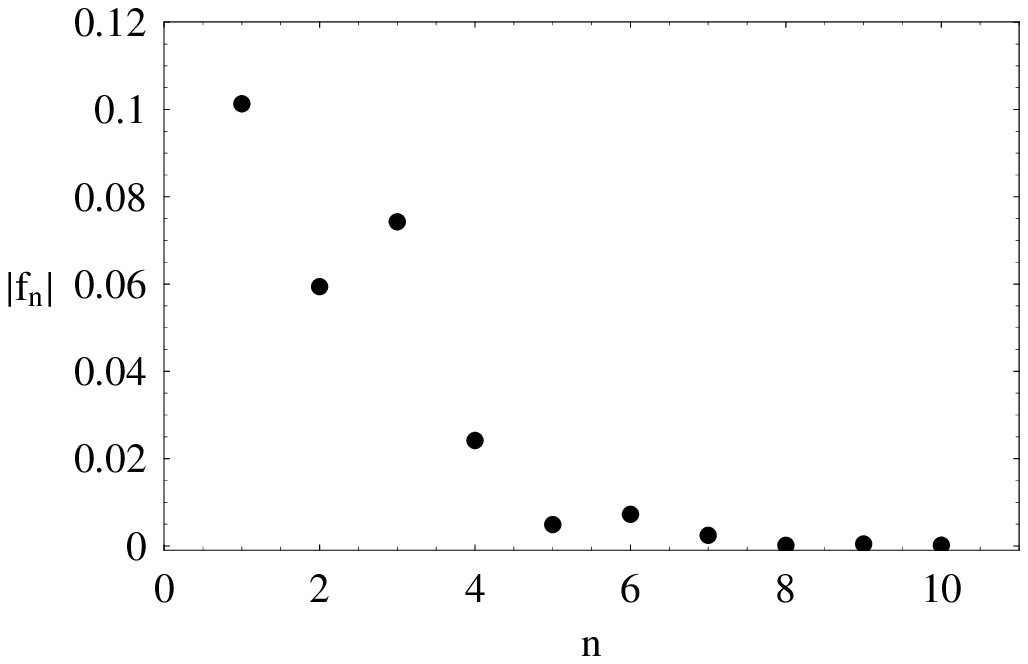}
   \put(-9.1,3.1){}
\put(-1.2,-.2){}
  \caption{}
\end{figure}
\newpage
\clearpage
\newpage
\setlength{\unitlength}{1cm}
\begin{figure}
 \includegraphics{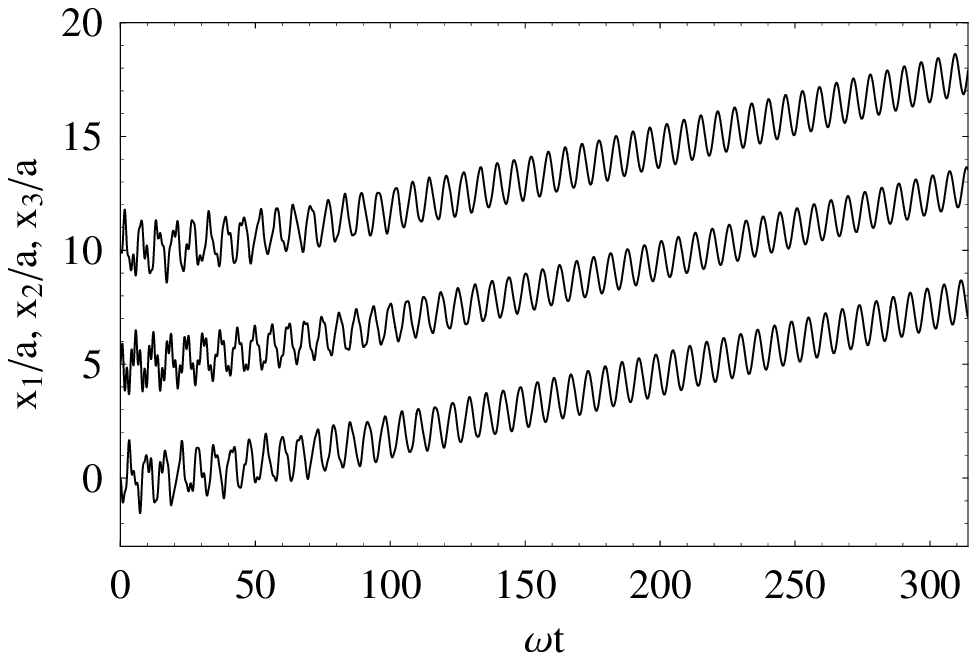}
   \put(-9.1,3.1){}
\put(-1.2,-.2){}
  \caption{}
\end{figure}
\newpage
\clearpage
\newpage
\setlength{\unitlength}{1cm}
\begin{figure}
 \includegraphics{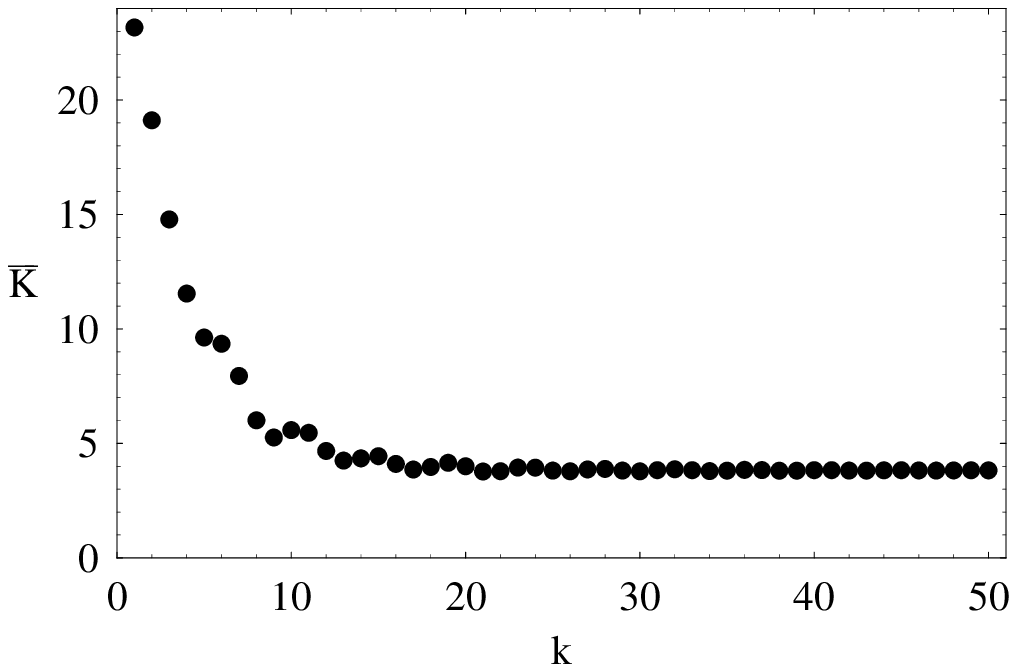}
   \put(-9.1,3.1){}
\put(-1.2,-.2){}
  \caption{}
\end{figure}
\newpage
\clearpage
\newpage
\setlength{\unitlength}{1cm}
\begin{figure}
 \includegraphics{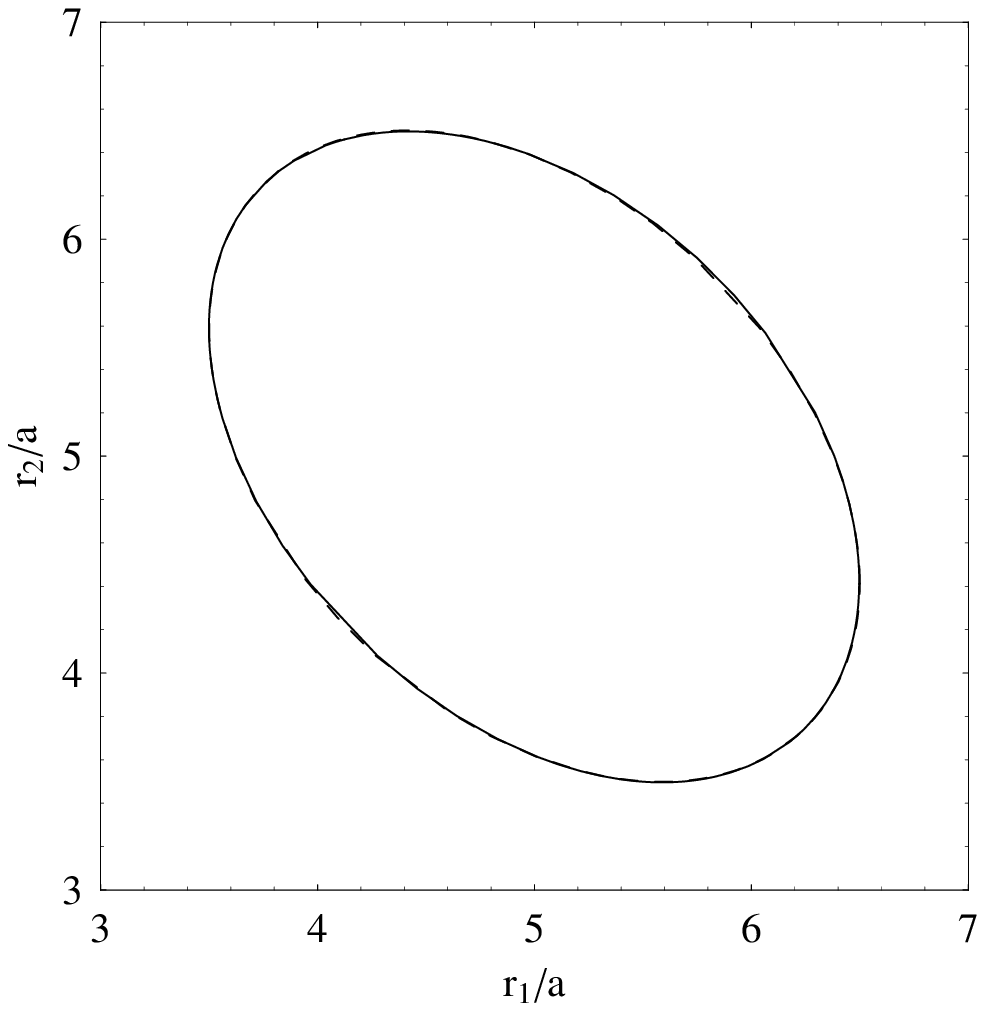}
   \put(-9.1,3.1){}
\put(-1.2,-.2){}
  \caption{}
\end{figure}
\newpage

\newpage

\begin{thebibliography}{99}

\bibitem{1}O. Lilienthal, {\it Der Vogelflug als Grundlage der Fliegekunst} (R. Gaertner's Verlagsbuchhandlung, Berlin, 1889).

\bibitem{2}
M. J. Lighthill, {\it Mathematical Biofluiddynamics} (SIAM , Philadelphia, 1975).

\bibitem{3}
S. Childress, {\it Mechanics of swimming and flying} (Cambridge University Press, Cambridge, 1981).

\bibitem{4}
C. J. Pennycuick, {\it Modelling the Flying Bird} (Academic Press, Burlington MA, 2008).
\bibitem{5}
H. Tennekes, {\it The Simple Science of Flight} (MIT Press, Cambridge MA, 2009).

\bibitem{6}
M. J. Lighthill, "Note on the swimming of slender fish", J. Fluid Mech. \vol{9}, 305 (1960).

\bibitem{7}
T. Y. Wu, "Swimming of a waving plate", J. Fluid Mech. \vol{10}, 321 (1961).

\bibitem{8}
T. Y. Wu, "On Theoretical Modeling of Aquatic and Aerial Animal Locomotion", Adv. Appl. Mech. \vol{38}, 291 (2001).

\bibitem{9}
T. Y. Wu, "Fish Swimming and Bird/Insect Flight", Annu. Rev. Fluid Mech. \vol{43}, 25 (2011).

\bibitem{10}
J. A. Sparenberg, "Survey of the mathematical theory of fish locomotion", J. Eng. Math. \vol{44}, 395 (2002).

\bibitem{11}
J. Gray, "Studies of animal locomotion", J. Expl. Biol. \vol{13}, 192 (1936).

\bibitem{12}
G. I. Taylor, "Analysis of the swimming of microscopic organisms", Proc. Roy. Soc. London A \vol{209}, 447 (1951).

\bibitem{13}
J. Lighthill, "Flagellar Hydrodynamics: The John von Neumann Lecture, 1975", SIAM Review \vol{18}, 161 (1976).

\bibitem{14}
E. Lauga and T. R. Powers, "The hydodynamics of swimming microorganisms", Rep. Prog. Phys. \vol{72}, 09660 (2009).

\bibitem{15}
B. U. Felderhof, "Swimming of an assembly of rigid spheres at low Reynolds number", Eur. Phys. J. E \vol{37}, 110 (2014).

\bibitem{16}
B. U. Felderhof, "Efficient swimming of an assembly of rigid spheres at low Reynolds number", arXiv1504.05794[physics.flu-dyn].

\bibitem{17}
J. Lighthill, "Large-amplitude elongated body theory of fish locomotion", Proc. R. Soc. Lond. B \vol{179}, 125 (1971).

\bibitem{18}
H.-B. Deng, Y.-Q. Xu, D.-D. Chen, H. Dai, J. Wian, and F.-B. Tian, "On numerical modeling of animal swimming and flight", Comput. Mech. \vol{52}, 1221 (2013).



\bibitem{19}
J. Happel and H. Brenner, {\it Low Reynolds number hydrodynamics} (Noordhoff, Leyden, 1973).

\bibitem{20}
J. Lighthill, {\it An Informal Introduction to Theoretical Fluid Mechanics} (Clarendon Press, Oxford, 1986).

\bibitem{21}
B. U. Felderhof, "Comment on: "Circular motion of asymmetric self-propelling particles"", Phys. Rev. Lett. \vol{113}, 029801 (2014).

\bibitem{22}
L. D. Landau and E. M. Lifshitz, {\it Fluid Mchanics} (Pergamon, Oxford, 1987).

\bibitem{23}
G. K. Batchelor, {\it An Introduction to Fluid Dynamics} (Cambridge University Press, Cambridge, 1967).

\bibitem{24}
B. U. Felderhof, "Virtual mass and drag in two-phase flow", J. Fluid Mech. \vol{225}, 177 (1991).

\bibitem{25}
R. Golestanian and A. J. Ajdari, "Analytic results for the three-sphere swimmer at low Reynolds number", Phys. Rev. E \vol{77}. 036308 (2008).

\bibitem{26}
B. U. Felderhof, "Optimization of flagellar swimming by a model sperm", arXiv:1412.3937.

\bibitem{27}
A. S. Sangani and A. Prosperetti, "Numerical simulation of the motion of particles at large Reynolds numbers" in {\it Particulate two-phase flow} ed. M. C. Roco (Butterworth-Heinemann, Boston, 1993).

\bibitem{28}
B. U. Felderhof, "The swimming of animalcules", Phys. Fluids \vol{18}, 063101 (2006).



\end{thebibliography}
\end{document}